\documentclass[10pt]{iopart}
\usepackage{graphicx}
\usepackage{iopams}

\newcommand{\half}{{\textstyle \frac{1}{2}}}
\newcommand{\third}{{\textstyle \frac{1}{3}}}
\newcommand{\fifth}{{\textstyle \frac{1}{5}}}
\newcommand{\twothirds}{{\textstyle \frac{2}{3}}}
\newcommand{\threehalves}{{\textstyle \frac{3}{2}}}
\newcommand{\fourthirds}{{\textstyle \frac{4}{3}}}
\newcommand{\twofifths}{{\textstyle \frac{2}{5}}}
\newcommand{\sixth}{{\textstyle \frac{1}{6}}}
\newcommand{\eighth}{{\textstyle \frac{1}{8}}}
\newcommand{\fourth}{{\textstyle \frac{1}{4}}}
\newcommand{\twelfth}{{\textstyle \frac{1}{12}}}
\newcommand{\threefourth}{{\textstyle \frac{3}{4}}}

\def\hbarit {{\mathchar'26\mkern-11muh}} 
\newcommand{\bfsfI}{\mbox{\sffamily\bfseries{I}}}

\newcommand{\bfsfF}{\mbox{\sffamily\bfseries{F}}}
\newcommand{\bfsfG}{\mbox{\sffamily\bfseries{G}}}

\newcommand{\bfsff}{\mbox{\sffamily\bfseries{f}}}

\eqnobysec 
\begin{document}
\jl{1} 

\title[Atomic decay]{Atomic decay near a quantized medium of absorbing scatterers}

\author{L G Suttorp and A J van Wonderen}

\address{Instituut voor Theoretische Fysica, Universiteit van Amsterdam,
Valckenierstraat 65, 1018 XE Amsterdam, The Netherlands}

\begin{abstract} 
The decay of an excited atom in the presence of a medium that both
scatters and absorbs radiation is studied with the help of a
quantum-electrodynamical model. The medium is represented by a half space filled
with a randomly distributed set of non-overlapping spheres, which consist
of a linear absorptive dielectric material. The absorption effects are
described by means of a quantized damped-polariton theory. It is found that
the effective susceptibility of the bulk does not fully account for the
medium-induced change in the atomic decay rate. In fact, surface effects
contribute to the modification of the decay properties as well. The
interplay of scattering and absorption in the total decay rate is
discussed.
\end{abstract}

\pacs{42.50.Nn, 42.25.Fx, 03.70.+k} 


\maketitle

\section{Introduction}\label{sec1}

The spontaneous-emission rate of an excited atom can be altered by the
atomic environment, as has been pointed out long ago \cite{P46}. For an
atom embedded in a uniform linear non-absorptive dielectric of infinite
extent the change in the emission rate has been obtained from quantum
electrodynamics \cite{D70,NA76}. For dense media local-field effects
have to be considered as well \cite{GL91,M95}. If the medium can absorb the
emitted photons, the analysis gets more complicated, since the loss
mechanism has to be treated in a quantum-mechanical context 
\cite{BHL92} - \cite{DKW00}.

Local-field effects do not play a role if a different geometry is
considered, with the atom situated outside a medium of finite (or
semi-infinite) extent. A well-known case is that of an atom in front of a
medium that fills a half-space. For this configuration the decay rate
depends on the distance between the atom and the medium
\cite{A75} - \cite{ICLS04}.

In all treatments mentioned so far the medium is structureless on the scale
of the wavelength corresponding to the atomic transition. As a consequence,
the medium properties are fully described by a susceptibility, which does
not vary appreciably on the scale of the wavelength. The picture changes
if the structure of the medium cannot be neglected, since scattering may
occur then as well. The total extinction in such a medium is driven by both
absorption and scattering. In practice, extinction by scattering in
material media is quite common. Impurities and defects both lead to
scattering effects that are difficult to avoid.

To study the interplay between the two types of extinction that may modify
atomic decay processes in the presence of a material medium, it is useful
to analyse a model in which both of these features occur
simultaneously. The model that we shall adopt in the following is that of a
medium consisting of non-overlapping spheres that are made of an absorptive
material. The spheres, which may move freely through the system, are
distributed randomly with a uniform average density. In a recent paper
\cite{YV2007} a similar model with a collection of spherical scatterers
consisting of non-absorptive material has been studied.

In order to describe the absorptive dielectric material of the spheres we
shall use the quantum-mechanical damped-polariton model. The central
quantity in this model is a space-dependent polarization density, which is
coupled to the electromagnetic field and to a bath of harmonic oscillators
accounting for absorption. The Hamiltonian of the damped-polariton model
can be diagonalized exactly, as has been shown both for the case of a
uniform dielectric \cite{HB92b} and for a dielectric with arbitrary
inhomogeneities \cite{SWo04}. 

To arrive at analytical results for the decay rate in the presence of a
medium with extinction due to both absorption and scattering we will adopt
several approximations. The density of the spherical scatterers will
supposed to be low, so that the medium is dilute. Furthermore, the size of
the spheres will be taken to be small as compared to the atomic
wavelength. Finally, the distance from the excited atom to the medium will
be chosen to be large compared to the wavelength.

The paper is organized as follows. In section 2 the properties of the model
and its diagonalization will be summarized. In section 3 the decay rate of
an excited atom in the presence of an arbitrarily inhomogeneous
damped-polariton dielectric will be derived from the basic Hamiltonian. The
decay rate is determined by the electromagnetic Green function, which
enters the description via a specific coefficient in the diagonalization
matrix. Since the medium consists of randomly distributed spheres the Green
function has to be averaged over all configurations in order to obtain the
physical decay rate. This averaging procedure will be discussed in section
4 and 5. As it turns out, the averaging process for a medium that fills a
finite region of space (or a half-space) should be carried out carefully,
since the boundaries give rise to specific surface contributions. Once
these surface effects have been evaluated for the specific case of a medium
filling a half-space, we can obtain the average decay rate of an excited
atom in the presence of such a medium. The change in the atomic
decay rate as a function of the distance between atom and medium will be
determined in section 6, and the interplay of absorption and scattering
processes will become clear. Some of the technical details of the
derivation are given in two appendices.

\section{Field quantization in the presence of an
  inhomogeneous absorbing dielectric medium}\label{sec2}

In the damped-polariton model the dielectric medium is described by a
polarization density, which interacts with the electromagnetic field
according to the standard minimal-coupling scheme. To account for
absorption effects a bath of harmonic oscillators with a continuous range
of eigenfrequencies is coupled to the polarization density in a bilinear
way. The Hamiltonian of the model is~\cite{HB92b,SWo04}
\begin{eqnarray} 
\fl H_d=\int \rmd {\bf r}\left[\frac{1}{2\varepsilon_0} \Pi^2
+ \frac{1}{2 \mu_0} (\bnabla\times {\bf A})^2 
+\frac{1}{2\rho} \, P^2+\half \,\rho\, {\omega}_0^2\, X^2
+\frac{1}{2\rho}\int_0^{\infty} \rmd \omega\, Q_{\omega}^2 \right. \nonumber\\
\fl \left.+\half\,\rho\int_0^{\infty} \rmd\omega\, \omega^2\, Y_{\omega}^2
+\frac{\alpha}{\rho}\,{\bf A}\cdot{\bf P}+\frac{\alpha^2}{2\rho}\, A^2
+\frac{1}{\rho}\int_0^{\infty}\rmd\omega\, v_{\omega}\, {\bf X}\cdot{\bf
Q}_{\omega}+\frac{1}{2\varepsilon_0}\, (\alpha {\bf X})_L^2\right]
\, .\label{2.1} 
\end{eqnarray}

 The transverse part of the electromagnetic field is determined by the
vector potential ${\bf A}({\bf r})$, for which the Coulomb gauge is
adopted. Its conjugate canonical momentum is ${\bf \Pi}({\bf r})$. The
linear dielectric, with a space-dependent density $\rho({\bf r})$, is
described by the harmonic displacement variable ${\bf X}({\bf r})$ and its
canonical momentum ${\bf P}({\bf r})$, with the associated eigenfrequency
$\omega_0({\bf r})$. The electromagnetic field is coupled to the dielectric
variables in the usual way. In terms of the polarization density
$-\alpha({\bf r}){\bf X}({\bf r})$, with a space-dependent coupling
parameter $\alpha({\bf r})>0$, the minimal-coupling scheme leads to an
electrostatic contribution involving $[\alpha({\bf r}) {\bf X}({\bf r})]_L$
and to a bilinear interaction term with ${\bf A}({\bf r})\cdot{\bf P}({\bf
r})$. The longitudinal part of a vector (or a tensor) is obtained by a
convolution with the longitudinal delta function $\bdelta_L ({\bf
r})=-\bnabla\bnabla(4\pi r)^{-1}$. Finally, damping is introduced in the
system by a continuum bath of harmonic oscillators with canonical variables
${\bf Y}_{\omega}({\bf r})$, ${\bf Q}_{\omega}({\bf r})$ and with
eigenfrequencies $\omega$. These bath oscillators are coupled to ${\bf
X}({\bf r})$ with a strength $v_{\omega}({\bf r})>0$.

The canonical variables obey the standard commutation relations:
\begin{eqnarray}
\fl \left[{\bf \Pi}({\bf r}),{\bf A}({\bf r}')\right]= 
-\rmi\,\hbarit\, \bdelta_T({\bf r}-{\bf r}') \qquad  \qquad 
\left[{\bf P}({\bf r}),{\bf X}({\bf r}')\right] =  
-\rmi\,\hbarit\, \bfsfI \, \delta({\bf r}-{\bf r}')\nonumber\\ 
\left[{\bf Q}_{\omega}({\bf r}),{\bf Y}_{\omega'}({\bf r}')\right] = 
-\rmi\,\hbarit\, \delta(\omega-\omega')\, \bfsfI \,\delta({\bf r}-{\bf
r}')
\label{2.2}
\end{eqnarray}
while all other commutators of the canonical variables vanish. Here
$\bfsfI$ is the three-dimensional unit tensor, while $ \bdelta_T({\bf
r})=\bfsfI\, \delta({\bf r}) -\bdelta_L({\bf r})$ is the transverse delta
function. The electric field operator ${\bf E}({\bf r})$ is the sum of a
transverse part depending on ${\bf \Pi}({\bf r})$ and a longitudinal part
that is proportional to the polarization density:
\begin{equation}
{\bf E}({\bf r})=-\frac{1}{\varepsilon_0}\, {\bf \Pi}({\bf
r})+\frac{1}{\varepsilon_0}\, [\alpha({\bf r}){\bf X}({\bf r})]_L\, .
\label{2.3}
\end{equation}

The Hamiltonian is quadratic in the canonical variables, and can be
diagonalized explicitly \cite{SWo04}:
\begin{equation}
H_d=\int \rmd {\bf r}\int_0^{\infty} \rmd\omega\, \hbarit\omega \, {\bf
C}^{\dagger}({\bf r},\omega)\cdot {\bf C}({\bf r},\omega)\label{2.4}
\end{equation}
where we omit a zero-point-energy term. The operators ${\bf C}({\bf
r},\omega)$ are annihilation operators, which (together with the associated
creation operators) satisfy the commutation relations:
\begin{equation}
\fl \left[{\bf C}({\bf r},\omega),{\bf C}^{\dagger}({\bf r}',\omega')\right]=
\delta(\omega-\omega')\, \bfsfI \, \delta({\bf r}-{\bf r}') \qquad \qquad
\left[{\bf C}({\bf r},\omega),{\bf C}({\bf r}',\omega')\right]=
0\, .
\label{2.5}
\end{equation}

Each canonical operator can be written as a linear combination of the
annihilation and creation operators. For instance, one has 
\begin{eqnarray}
{\bf A}({\bf r})=\int \rmd{\bf r}'\int_0^{\infty} \rmd\omega \, \bfsff_A({\bf
r},{\bf r}',\omega)\cdot{\bf C}({\bf r}',\omega) +{\rm h.c.}
\label{2.6}\\ 
{\bf E}({\bf r})=\int \rmd{\bf r}'\int_0^{\infty} \rmd\omega \,
\bfsff_E({\bf r},{\bf r}',\omega)\cdot{\bf C}({\bf r}',\omega) +{\rm
h.c.} \label{2.7} 
\end{eqnarray} 
with the tensorial coefficients $\bfsff_A$ and $\bfsff_E$. Similar expressions
can be written for the other canonical variables. Since the vector
potential is transverse, the coefficient $\bfsff_A$ is transverse in ${\bf
r}$.

In order to derive explicit expressions for the coefficients one may use a
method due to Fano \cite{F61}. It amounts to evaluating the commmutator of
$C({\bf r},\omega)$ with the Hamiltonian $H_d$ in two different ways. On
the one hand this commutator follows from \eref{2.4} and \eref{2.5} as
$[C({\bf r},\omega),H_d]=\hbarit\, \omega\, C({\bf r},\omega)$, and on the
other hand it may be evaluated by first writing $C({\bf r},\omega)$ as a
linear combination of the canonical variables, subsequently inserting
\eref{2.1} and finally employing \eref{2.2}. Upon solving the linear
equations that follow by comparing the results of these two approaches, one
arrives at explicit expressions for the coefficients in terms of the
tensorial Green function $\bfsfG$ of the system \cite{SWo04}. The latter is
defined as the solution of the standard equation
\begin{equation}
\fl -\bnabla\times [\bnabla\times \bfsfG ({\bf r},{\bf
 r}',\omega+\rmi 0)]+\frac{\omega^2}{c^2}\, \left[1+\chi({\bf
 r},\omega+\rmi 0)\right]\, 
\bfsfG({\bf r},{\bf r}',\omega+\rmi 0)=\bfsfI\, \delta({\bf r}-{\bf r}')
 \label{2.8}
\end{equation}
with $\omega+\rmi 0$ in the upper half of the complex plane and
infinitesimally close to the positive real axis. In the course of the
diagonalization process the frequency- and position-dependent
susceptibility $\chi$ is found as
\begin{equation}
\chi({\bf r},\omega+\rmi 0)=-\frac{\alpha^2}{\varepsilon_0\rho}\, 
\left[\omega^2-\omega_0^2-\frac{1}{\rho^2}
\int_0^{\infty} \rmd\omega'\, \frac{{\omega'}^2\, v_{\omega'}^2}
{(\omega+\rmi 0)^2-{\omega'}^2}\right]^{-1}\, . \label{2.9}
\end{equation}
The tensorial Green function satisfies the reciprocity relation
\begin{equation}
\tilde{\bfsfG}({\bf r},{\bf
r}',\omega+\rmi 0)=\bfsfG({\bf r}',{\bf r},\omega+\rmi 0)
\label{2.10}
\end{equation}
where the tilde denotes the tensor transpose. In terms of the above
tensorial Green function, coefficients \eref{2.6} and \eref{2.7} are
given as 
\begin{eqnarray}
\bfsff_E({\bf r},{\bf r}',\omega)=-i\, \frac{\omega^2}{c^2}
\left(\frac{\hbarit \, {\rm Im}\,\chi({\bf
r}',\omega+i0)}{\pi\varepsilon_0}\right)^{1/2}\, 
\bfsfG({\bf r},{\bf r}',\omega+\rmi 0) \label{2.11}\\
\bfsff_A({\bf r},{\bf r}',\omega)=-\frac{i}{\omega}\, 
[\bfsff_E({\bf r},{\bf r}',\omega)]_T \, .\label{2.12}
\end{eqnarray} 
These expressions follow from the results presented in \cite{SWo04}. When
they are substituted in \eref{2.6} and \eref{2.7}, they lead to expressions
for the vector potential and the electric field that agree with those
postulated in a phenomenological quantization scheme \cite{SKWB99},
\cite{SKW99a}, \cite{DKW00}.

\section{Decay of an excited atom in the presence of an inhomogeneous absorbing
dielectric}\label{sec3} 

When a neutral atom at a fixed position is present as well, the total
Hamiltonian $H$ of the system is given by the sum $H_d+H_a+H_i$ of the
damped-polariton Hamiltonian \eref{2.1}, the atomic Hamiltonian
\begin{equation}
H_{a}=\sum_i \frac{p_i^2}{2m}+\sum_{i\neq j}
\frac{e^2}{8\pi\varepsilon_0|{{\bf r}_i-{\bf r}_j}|}
-\sum_i \frac{Ze^2}{4\pi\varepsilon_0|{\bf r}_i-{\bf r}_a|}
\label{3.1}
\end{equation}
(with $Z$ the atomic number, ${\bf r}_a$ the fixed position of the nucleus
and ${\bf r}_i\, , \, {\bf p}_i$ the positions and momenta of the
electrons) and the interaction Hamiltonian $H_i$, which follows from the
usual minimal-coupling scheme as
\begin{equation} 
H_i=\int \rmd{\bf r} [\rho_a({\bf r})\, \varphi({\bf r})-{\bf J}_a({\bf r})
\cdot {\bf A}({\bf r})] \, . \label{3.2}
\end{equation}
Here $\rho_a({\bf r})=e\sum_i[\delta({\bf r}-{\bf r}_a)-\delta({\bf r}-{\bf
r}_i)]$ and ${\bf J}_a({\bf r})=-\half e\sum_i\{{\bf p}_i/m,\delta({\bf
r}-{\bf r}_i)\}$ are the local atomic charge and current densities, with
curly brackets denoting the anticommutator. Furthermore, $\varphi({\bf r})$
is the scalar potential due to the polarization of the dielectric. Its
gradient is given by
\begin{equation} 
\fl \bnabla\varphi({\bf r})=-\frac{1}{\varepsilon_0}\, [\alpha({\bf r})\, {\bf
X}({\bf r})]_L=-{\bf E}({\bf r})-\frac{1}{\varepsilon_0}\, {\bf \Pi}({\bf
r})=-{\bf E}({\bf r})-\dot{\bf A}({\bf r})
\label{3.3} 
\end{equation}
with the time derivative given as $\dot{\bf A}=(\rmi/\hbarit)[H,{\bf
A}]$. In writing \eref{3.2} we have assumed that local-field effects are
negligible.

We assume that at the initial time $t=0$, the atom is prepared in an excited
state $|e\rangle$, while the dielectric medium (including the bath) and the
electromagnetic field are in the ground state $|0\rangle$ of the
Hamiltonian \eref{2.1} or \eref{2.4}, i.e.\ in the state that is
annihilated by all operators ${\bf C}({\bf r},\omega)$. The atom will decay
to its ground state $|g\rangle$ with a time-dependent decay rate
$\Gamma(t)$. This rate follows from perturbation theory in leading order as
\begin{equation}
\Gamma(t)=\frac{1}{\hbarit^2}\sum_f \int_0^t \rmd t'\,
\rme^{(\rmi/\hbarit)(E_f-E_i)t'}\, 
|\langle i|H_i|f\rangle|^2+{\rm c.c.}
\label{3.4}
\end{equation}
with $|i\rangle$ and $|f\rangle$ the initial and final states of the
total system, with energies $E_i$ and $E_f$, respectively.

Upon taking the matrix element of \eref{3.2}, using charge conservation in
the form $\bnabla\cdot{\bf J}_a({\bf r})=-(\rmi/\hbarit)[H_a,\rho_a({\bf
r})]$, carrying out a partial integration and substituting \eref{3.3} with
\eref{2.6}-\eref{2.7}, we may rewrite the time-dependent decay rate as
\begin{eqnarray}
\fl \Gamma(t)=\frac{1}{\hbarit^2  \omega_a^2}
\int_0^t \rmd t' \int \rmd{\bf r} \int \rmd{\bf r}'
\int \rmd{\bf r}'' \int_0^\infty \rmd\omega\,
\rme^{\rmi(\omega-\omega_a)t'}\nonumber\\
\langle e|{\bf J}_a({\bf r}')|g\rangle \cdot
\left[(\omega_a-\omega)\, \bfsff_A({\bf r}',{\bf r},\omega)-\rmi\,  
\bfsff_E({\bf r}',{\bf r},\omega)\right]\cdot\nonumber\\
\cdot \left[(\omega_a-\omega)\, \tilde{\bfsff}_A^\ast({\bf r}'',{\bf r},\omega)+\rmi\,  
\tilde{\bfsff}_E^\ast({\bf r}'',{\bf r},\omega)\right]\cdot 
\langle g|{\bf J}_a({\bf r}'')|e\rangle
+{\rm c.c.}\label{3.5}
\end{eqnarray} 
where $\hbarit \omega_a$ is the difference between the energies of the excited
and the ground state of the atom.

For large values of $t$ the decay rate becomes independent of
time. Carrying out the integrals over $t'$ and $\omega$ one finds that the 
coefficient $\bfsff_A$ drops out. As a result we get
\begin{equation}
\fl \Gamma=\frac{2\pi}{\hbarit^2 \omega_a^2}
\int \rmd{\bf r} \int \rmd{\bf r}' \int \rmd{\bf r}'' \, 
\langle e|{\bf J}_a({\bf r}')|g\rangle \cdot
\bfsff_E({\bf r}',{\bf r},\omega_a) 
\cdot\tilde{\bfsff}_E^\ast({\bf r}'',{\bf r},\omega_a)\cdot 
\langle g|{\bf J}_a({\bf r}'')|e\rangle\, .
\label{3.6}
\end{equation}
Inserting \eref{2.11} we obtain
\begin{equation}
\fl \Gamma=
\frac{2\omega_a^2}{\varepsilon_0 \hbarit c^4}
\int \rmd{\bf r}' \int \rmd{\bf r}'' \, 
\langle e|{\bf J}_a({\bf r}')|g\rangle \cdot
\bfsfF({\bf r}',{\bf r}'',\omega_a)\cdot 
\langle g|{\bf J}_a({\bf r}'')|e\rangle
\label{3.7}
\end{equation}
with the abbreviation
\begin{equation}
\fl \bfsfF({\bf r}',{\bf r}'',\omega)=
\int \rmd{\bf r}\, 
\bfsfG({\bf r}',{\bf r},\omega+\rmi 0)
\cdot\tilde{\bfsfG}^\ast({\bf r}'',{\bf r},\omega+\rmi 0)\, 
{\rm Im}\, \chi({\bf r},\omega+\rmi 0)\, .
\label{3.8}
\end{equation}
With the help of the differential equation \eref{2.8} and the reciprocity
relation \eref{2.10} one may rewrite $\bfsfF({\bf r},{\bf r}',\omega)$ as
$-(c/\omega)^2\, {\rm Im}\, \bfsfG({\bf r},{\bf r}',\omega+\rmi 0)$, so
that the decay rate gets the final form
\begin{equation}
\fl \Gamma= -\frac{2}{\varepsilon_0 \hbarit c^2} 
\int \rmd{\bf r}' \int \rmd{\bf r}'' \, 
\langle e|{\bf J}_a({\bf r}')|g\rangle \cdot
{\rm Im}\, \bfsfG({\bf r}',{\bf r}'',\omega_a+\rmi 0)\cdot 
\langle g|{\bf J}_a({\bf r}'')|e\rangle
\, .\label{3.9}
\end{equation}
In the electric-dipole approximation the matrix element $\langle e|{\bf
J}_a({\bf r})|g\rangle$ is replaced by its localized form $\delta({\bf
r}-{\bf r}_a)\, \int \rmd{\bf r}\langle e|{\bf J}_a({\bf r})|g\rangle=i\, 
\delta({\bf r}-{\bf r}_a)\, \omega_a \langle e |\bmu|g\rangle$, with
$\bmu=-e\sum_i({\bf r}_i-{\bf r}_a)$ the electric dipole moment.
In that approximation the decay rate reads
\begin{equation}
\Gamma= -\frac{2\omega_a^2}{\varepsilon_0 \hbarit c^2} 
\,\langle e|\bmu|g\rangle \cdot
{\rm Im}\, \bfsfG({\bf r}_a,{\bf r}_a,\omega_a+\rmi 0)\cdot 
\langle g|\bmu|e\rangle \, .
\label{3.10}
\end{equation}
This expression for the decay rate, which is valid for an excited atom
in the presence of an absorptive dielectric with arbitrary inhomogeneities, can
be obtained as well by invoking the fluctuation-dissipation theorem
\cite{BHLM96,SKW99a}. The above derivation shows how it follows from the
explicit diagonalization of the inhomogeneous damped-polariton model in a
rigorous way.

\section{Medium of absorbing spherical scatterers}\label{sec4}

Let us consider a medium of non-overlapping spheres of absorptive
material. It is an inhomogeneous dielectric that may be described by 
Hamiltonian \eref{2.1}. The susceptibility \eref{2.9} has a constant value
within the spheres and vanishes outside, so that it may be written as
$\chi({\bf r}, \omega+\rmi 0)=\chi(\omega+\rmi 0)\, f({\bf r})$. If the
radius of the spheres is $a$ and the centre of the sphere $i$ is located at
${\bf r}_i$ (with $|{\bf r}_i-{\bf r}_j|\geq 2a$ for $i\neq j$ so as to
avoid overlap), the function $f({\bf r})$ equals $\sum_i \theta(a-|{\bf
r}-{\bf r}_i|)$, with the step function $\theta(x)$ equal to 1 for $x>0$
and 0 elsewhere.

We are interested in the decay of an excited atom in the presence of such a
medium of absorptive spheres. Since the spheres may move the effective
decay rate follows from \eref{3.9} or \eref{3.10} by averaging over the
positions of the centers of the spheres. Hence, we have to find an
expression for the (imaginary part of the) average Green function.

The differential equation \eref{2.8} for the Green function $\bfsfG({\bf
r},{\bf r}',\omega+\rmi 0)$ is equivalent to an integral equation that
relates $\bfsfG$ to the vacuum Green function
$\bfsfG_0$:
\begin{equation}
\fl \bfsfG({\bf r},{\bf r}', z)=\bfsfG_0({\bf r},{\bf r}', z)-
\frac{z^2}{c^2}\, \chi(z)\int \rmd{\bf r}''\, f({\bf r}'')\, 
\bfsfG_0({\bf r},{\bf r}'', z)\cdot \bfsfG({\bf r}'',{\bf r}', z)
\label{4.1}  
\end{equation}
with the frequency variable $z=\omega+\rmi 0$. The vacuum Green function is
the solution of \eref{2.8} with $\chi({\bf r},\omega+\rmi 0)=0$.
Iterating \eref{4.1} we get a series of terms, which up to second order
in the susceptibility reads
\begin{eqnarray}
\fl \bfsfG({\bf r},{\bf r}', z)=\bfsfG_0({\bf r},{\bf r}', z)-
\frac{z^2}{c^2}\, \chi(z)\int \rmd{\bf r}''\, f({\bf r}'')\, 
\bfsfG_0({\bf r},{\bf r}'', z)\cdot \bfsfG_0({\bf r}'',{\bf r}',z)\nonumber\\
\fl +\frac{z^4}{c^4}\, [\chi(z)]^2\int \rmd {\bf r}''\int\rmd {\bf r}'''\, 
f({\bf r}'')\, f({\bf r}''')  \, \bfsfG_0({\bf r},{\bf r}'', z)\cdot 
\bfsfG_0({\bf r}'',{\bf r}''', z)\cdot 
\bfsfG_0({\bf r}''',{\bf r}', z)
+ \ldots\nonumber \\
\mbox{}
\label{4.2}  
\end{eqnarray}

When averaging both sides of this equation over the positions of the
centers of the spheres, we need expressions for the averages $\langle f({\bf
r})\rangle$ and $\langle f({\bf r})\, f({\bf r}')\rangle$. When the centers
of the spheres are uniformly distributed with the density $n$, these averages
have the form
\begin{eqnarray}
\langle f({\bf r})\rangle= n\, v_0 \label{4.3}\\
\langle f({\bf r})\, f({\bf r}')\rangle= 
n\int \rmd{\bf r}''\, \theta(a-|{\bf r}-{\bf r}''|)\,\theta(a-|{\bf r}'-{\bf
r}''|)\nonumber\\
+n^2 \int \rmd{\bf r}''\int d{\bf r}'''\, \theta(a-|{\bf r}-{\bf r}''|)\,
\theta(a-|{\bf r}'-{\bf r}'''|)\, g({\bf r}'',{\bf r}''') \label{4.4}
\end{eqnarray}
with $v_0=4\pi a^3/3$ being the volume of the spheres and $g({\bf r},{\bf r}')$
being the pair correlation function. If the spheres are dilutely distributed, the
correlations may be neglected, so that $g$ can be replaced by
unity. Upon carrying out the geometrical integrals expression \eref{4.4} becomes
\begin{equation}
\fl \langle f({\bf r})\, f({\bf r}')\rangle=  
n\, \left[v_0-\pi a^2\, |{\bf r}-{\bf r}'|+\twelfth\,\pi\,|{\bf r}-{\bf
r}'|^3\right]\, \theta(2\, a- |{\bf r}-{\bf r}'|) 
+n^2\, v_0^2 \, .\label{4.5}
\end{equation}

Inserting the above averages in the iterated integral equation \eref{4.2},
we get up to second order in the susceptibility:
\begin{eqnarray}
\fl \langle \bfsfG({\bf r},{\bf r}', z)\rangle =\bfsfG_0({\bf r},{\bf r}',z)-
\frac{z^2}{c^2}\, n\, v_0\, \chi(z) \int d{\bf r}'' \int d{\bf r}'''\,
\bfsfG_0({\bf r},{\bf r}'',z)
\cdot
\nonumber\\
\cdot\left[ \bfsfI \, \delta({\bf r}''-{\bf r}''')
-\frac{z^2}{c^2}\, \chi(z)\, n\, v_0\, \bfsfG_0({\bf r}'',{\bf r}''',z)\right.\nonumber\\
\left.-\frac{z^2}{c^2}\, \chi(z)\, c(|{\bf r}''-{\bf r}'''|)\, 
\bfsfG_0({\bf r}'',{\bf r}''',z)
\right] \cdot \bfsfG_0({\bf r}''',{\bf r}',z) + \ldots 
\label{4.6}
\end{eqnarray} 
with the abbreviation 
\begin{equation}
c(r)=\left(1-\frac{3r}{4a}+\frac{r^3}{16a^3}\right)\, \theta(2\, a-r)\, .
\label{4.7}
\end{equation}
The right-hand side is the iterated solution of the integral equation
\begin{eqnarray}
\fl \langle \bfsfG({\bf r},{\bf r}', z)\rangle =\bfsfG_0({\bf r},{\bf r}',z)
-\frac{z^2}{c^2}\int d{\bf r}'' \int d{\bf r}'''\,
\bfsfG_0({\bf r},{\bf r}'',z)
\cdot \bchi_e({\bf r}'',{\bf r}''',z)\cdot
\langle \bfsfG({\bf r}''',{\bf r}', z)\rangle
\nonumber\\
\mbox{}
\label{4.8}
\end{eqnarray} 
again up to second order in $\chi(z)$. The effective susceptibility tensor
is given as
\begin{equation}
\fl \bchi_e({\bf r},{\bf r}', z) =n\, v_0\, \chi(z)\, 
\left[  \bfsfI \, \delta({\bf r}-{\bf r}')
-\frac{z^2}{c^2}\, \chi(z)\, c(|{\bf r}-{\bf r}'|)\, 
\bfsfG_0({\bf r},{\bf r}',z)\right] \, .
\label{4.9}
\end{equation} 
It should be noted that the term proportional to $n^2\, [\chi(z)]^2$ in
\eref{4.6} results upon iterating \eref{4.8} up to second order in $\bchi_e$.

The effective susceptibility \eref{4.9} is non-local with a range equal to
$2a$. The Green functions in the integrand of the last term of \eref{4.8}
do not change appreciably over that range, when $a$ is small compared to
$c/\omega$ (for $z=\omega+\rmi 0$), and to $|{\bf r}-{\bf r}''|$ and $|{\bf
r}'-{\bf r}'''|$. The first of these conditions can easily be met for
spheres that are sufficiently small. In fact, we shall use \eref{4.8} for
$\omega$ equal to the atomic transition frequency, so that $c/\omega$
equals the transition wavelength. In contrast, the last two conditions are
fulfilled only when ${\bf r}''$ and ${\bf r}'''$ in \eref{4.8} are
sufficiently far from the fixed positions ${\bf r}$ and ${\bf r}'$. Since
the integrations in \eref{4.8} extend over all parts of space accessible to
the spheres, it is not obvious that the two conditions can be fulfilled. We
will postpone a discussion of this point to the end of the section.

When all three conditions mentioned above are satisfied, one may replace
$\bchi_e$ by its localized version. Quite generally the localized version
of a function $F({\bf r})$ that is of short range and centred around the
origin can be written as a series expansion of which the first few terms
are
\begin{eqnarray}
\fl F({\bf r})=\delta({\bf r}) \int \rmd{\bf r}'\, F({\bf r}')-
[\bnabla\, \delta({\bf r})]\cdot  \int \rmd{\bf r}'\, {\bf r}'\,
F({\bf r}')+\half [\bnabla\bnabla\, \delta({\bf r})]\,  :
\int \rmd{\bf r}'\, {\bf r}'{\bf r}'\, F({\bf r}')+\ldots\nonumber\\
\label{4.10}
\end{eqnarray} 
We can evaluate the first few moments of $F({\bf r})=c(|{\bf r}|)\, \bfsfG_0({\bf
r},0,\omega+\rmi 0)$ for small values of $q=\omega a/c$ by employing the
expression \cite{W89,VCL98}
\begin{eqnarray}
\fl \bfsfG_0({\bf r},0,z)=-\frac{1}{4\pi r}\, \left(\bfsfI-\frac{{\bf rr}}{r^2}\right)\,
\rme^{\rmi zr/c}
+{\cal P}\frac{1}{4\pi r}
\left(-\rmi\, \frac{c}{zr}+\frac{c^2}{z^2r^2}\right)\, 
\left(\bfsfI-3\frac{{\bf rr}}{r^2}\right)\, \rme^{\rmi zr/c}\nonumber\\
+\frac{c^2}{3 z^2}\, \delta({\bf r})\, \bfsfI 
\label{4.11}
\end{eqnarray}
for the vacuum Green function (with $z$ in the upper part of the complex
plane). The principal-value sign denotes the exclusion of an infinitely
small spherical volume in subsequent integrations.  As a result we obtain
the following for the localized form of $\bchi_e$:
\begin{eqnarray}
\fl \bchi_e({\bf r},{\bf r}',z)=n\, v_0 \chi(z)\,
\left[ 1-\chi(z)\, \left(\frac{1}{3}-\frac{4}{15}\, q^2-\frac{2\rmi}{9}\,
q^3\right)\right]\, \bfsfI\, \delta({\bf r}-{\bf r}')\nonumber\\
-n\, v_0\, [\chi(z)]^2\,\frac{2}{75}\, a^2\, (\bfsfI\, \Delta
-3\bnabla\bnabla)\, \delta({\bf r}-{\bf r}')\, .  \label{4.12}
\end{eqnarray}
Here we have used the integral identities $\int d\Omega\, r_i\, r_j/r^2
=(4\pi/3)\, \delta_{ij}$ and $\int d\Omega\, r_i\, r_j\, r_k\, r_l/r^4
=(4\pi/15)\, (\delta_{ij}\, \delta_{kl}+\delta_{ik}\,
\delta_{jl}+\delta_{il}\, \delta_{jk})$, with $r_i$ being the cartesian
components of the position vector ${\bf r}$ and $d\Omega$ being an element of
solid angle in the direction of ${\bf r}$.

When the localized form \eref{4.12} is inserted in \eref{4.8} and a
partial integration with respect to ${\bf r}''$ is carried out, we can
employ the identity 
\begin{equation}
(\bfsfI\, \Delta - 3\, \bnabla\bnabla)\cdot
\bfsfG_0({\bf r},{\bf r}',z)=-\frac{z^2}{c^2}\, 
\bfsfG_0({\bf r},{\bf r}',z) \label{4.13}
\end{equation}
for ${\bf r}\neq {\bf r}'$, as follows from the differential equation
\eref{2.8} (with $\chi=0$) for the vacuum Green function. It should be
noted that the second operator between the brackets in \eref{4.13} does not
contribute for ${\bf r}\neq {\bf r}'$, as is obvious from the form of the
differential equation. There is no need to discuss the form that
\eref{4.13} may take for ${\bf r}={\bf r}'$, since the localized form of
the effective susceptibility can be used only when the arguments of the
Green function are sufficiently far apart, as we have seen above.

When the identity \eref{4.13} is taken into account, the localized form
\eref{4.12} of the effective susceptibility may be rewritten as
\begin{eqnarray}
\fl \bchi_e({\bf r},{\bf r}',z)=n\, v_0 \chi(z)\,
\left[ 1-\chi(z)\, \left(\frac{1}{3}-\frac{22}{75}\, q^2-\frac{2\rmi}{9}\,
q^3\right)\right]\, \bfsfI\, \delta({\bf r}-{\bf r}')
\equiv\chi_e(z)\, \bfsfI\, \delta({\bf r}-{\bf r}') \, .  \nonumber\\
\mbox{}\label{4.14}
\end{eqnarray}
This localized form of the susceptibility has to be inserted in the integral
equation \eref{4.8}. Its solution up to second order in $\chi(z)$ and up
to first order in $n$ reads
\begin{equation}
\fl \langle \bfsfG({\bf r},{\bf r}',z)\rangle=\bfsfG_0({\bf r},{\bf
r}',z)
-\frac{z^2}{c^2}\, \chi_e(z) \int d{\bf r}'' \, 
\bfsfG_0({\bf r},{\bf r}'',z) \cdot \bfsfG_0({\bf r}'',{\bf r}',z).
\label{4.15}
\end{equation}

In principle, this expression for the average Green function, with ${\bf
r}={\bf r}'={\bf r}_a$ and $z=\omega_a+\rmi 0$, could be used to evaluate
the right-hand side of \eref{3.10}. When the medium of scattering and
absorbing spheres is infinitely large, the atom is necessarily embedded in
the medium. Since the integrals in \eref{4.8} have to be taken over all
space in that case, the integration variables ${\bf r}''$ and ${\bf r}'''$
can coincide with ${\bf r}_a$, so that the validity of the localized form
\eref{4.14} of the effective susceptibility is not guaranteed. This comes
as no surprise, as local-field effects have to be taken into account in
such a situation. We have to conclude that \eref{4.15} cannot be used as
such to determine the average decay rate of an excited atom in an infinite
medium of absorptive spheres. However, one is often interested in a
scattering medium of finite extent, in which the centers of the spheres are
confined to a volume $V$ (with $V/a^3\gg 1$), while the excited atom is
situated outside $V$. This configuration will be considered in the next
section. The localized effective susceptibility is a useful concept in that
case and the expression \eref{4.15} for the average Green function can be
employed, albeit after a suitable modification.

\section{Finite media and surface effects}\label{sec5}
For a finite medium the expression for the average Green function has to be
modified so as to include finite-volume effects. We start again from
\eref{4.2} and take the average over the positions of the centers of the
spheres, which must be inside $V$. In lowest order of the susceptibility
one encounters the average $\langle f({\bf r})\rangle$, which for a
uniformly distributed set of spheres with centers in $V$ is given by
\begin{eqnarray} 
\fl \langle f({\bf r})\rangle= n\int^V d{\bf r}'\, 
\theta(a-|{\bf r}-{\bf r}'|)= 
n\, v_0\, \theta_V({\bf r}) +n\left[\int^V d{\bf r}'\,
\theta(a-|{\bf r}-{\bf r}'|)-v_0\, \theta_V({\bf r})\right]\nonumber\\
\label{5.1} 
\end{eqnarray} 
instead of \eref{4.3}. The step function $\theta_V({\bf r})$ equals 1 for
${\bf r}$ inside $V$ and vanishes elsewhere. The expression between square
brackets differs from 0 only for positions ${\bf r}$ that are close to the
surface of $V$, at a distance less than $a$. Assuming the surface to be
approximately flat on that scale, one may write ${\bf r}$ as ${\bf r}_s+h\,
{\bf n}$, with ${\bf r}_s$ a position vector at the surface and ${\bf n}$ a
unit vector normal to the surface at ${\bf r}_s$ and pointing outwards. In
that notation one finds
\begin{equation} 
\langle f({\bf r})\rangle=n\, v_0\, \theta_V({\bf r}) 
+n\left[\half\, v_0\, \varepsilon(h)-\pi\, a^2\, h+\third \, \pi\,
h^3\right]\,\theta(a-|h|) 
\label{5.2}
\end{equation}
with $\varepsilon(x)=\theta(x)-\theta(-x)$.  As a consequence the
contribution of $\langle\bfsfG\rangle$ that is linear in the susceptibility
$\chi(z)$ gets the form
\begin{eqnarray}
 -\frac{z^2}{c^2}\, n\, v_0 \, \chi(z) \int^V d{\bf r}'' \, 
\bfsfG_0({\bf r},{\bf r}'',z) \cdot \bfsfG_0({\bf r}'',{\bf r}',z)\nonumber\\
 -\frac{z^2}{c^2}\, n \,\chi(z)\int^S dS''\int^a_{-a}dh''\, 
\left[\half\, v_0\,\varepsilon(h'')-\pi\, a^2\, h''+\third \, \pi\,
h''^3\right]\nonumber\\
\times \bfsfG_0({\bf r},{\bf r}''_s+h''\,{\bf n}'',z) \cdot 
\bfsfG_0({\bf r}''_s+h''\,{\bf n}'',{\bf r}',z)\, .
\label{5.3}
\end{eqnarray}
The first term is the bulk contribution. It has the same form as the term
linearly dependent on $\chi(z)$ in \eref{4.15}, with the integration
extended over $V$ only.  The second term is the surface contribution. Here
$dS''$ is a surface element at ${\bf r}''_s$, with a local normal unit
vector ${\bf n}''$.

The surface contribution in \eref{5.3} may be evaluated as follows. Let us
assume that both ${\bf r}$ and ${\bf r}'$ are far outside the volume, so
that both $|{\bf r}-{\bf r}''_s|$ and $|{\bf r}'-{\bf r}''_s|$ are much
larger than the wavelength (which itself is much larger than the radius of
the spheres). In that case the dependence of the Green functions $\bfsfG_0$
on $h''$ is determined by a phase factor, as follows from \eref{4.11}. As a
consequence one may write
\begin{eqnarray}
\bfsfG_0({\bf r},{\bf r}''_s+h''\,{\bf n}'',z) \cdot 
\bfsfG_0({\bf r}''_s+h''\,{\bf n}'',{\bf r}',z)=\nonumber\\
=\bfsfG_0({\bf r},{\bf r}''_s,z) \cdot 
\bfsfG_0({\bf r}''_s,{\bf r}',z)\, \rme^{-\rmi\, z\, h''\, {\bf
n}''\cdot ({\bf e}_s+{\bf e}'_s)/c}
\label{5.4}
\end{eqnarray}
with ${\bf e}_s$ and ${\bf e}'_s$ being unit vectors in the direction ${\bf r}-{\bf
r}''_s$ and ${\bf r}'-{\bf r}''_s$, respectively. For
$z=\omega+\rmi\, 0$ and $q=\omega a/c\ll 1$, as before, the exponential can
be expanded. Subsequently, upon evaluating the integral over $h''$ in
\eref{5.3} we arrive at a surface contribution of the form
\begin{equation}
\fl  \rmi\,\frac{z^2}{10c^2}\, n\, v_0\,\chi(z)\, a\, q\int^S dS''\,  
{\bf n}''\cdot ({\bf e}_s+{\bf e}'_s)\, 
\bfsfG_0({\bf r},{\bf r}''_s,z) \cdot 
\bfsfG_0({\bf r}''_s,{\bf r}',z)\, .
\label{5.5}
\end{equation}

With the use of Gauss's theorem, the surface integral can be transformed to a
volume integral. Since ${\bf r}$ and ${\bf r}'$ are both far from the
surface, the ensuing differentiation operator can be taken to act on the
phase factors in the Green functions only. Carrying out these
differentiations we finally arrive at the following expression for the
surface contribution in $\langle\bfsfG\rangle$ that is linear in the
susceptibility:
\begin{eqnarray}
\fl  \frac{z^2}{5c^2}\, n\, v_0\, \chi(z) \, q^2\int^V d{\bf r}'' \, 
(1+{\bf e}\cdot{\bf e}') \, 
\bfsfG_0({\bf r},{\bf r}'',z) \cdot \bfsfG_0({\bf r}'',{\bf r}',z)
\label{5.6}
\end{eqnarray}
with ${\bf e}$ being a unit vector in the direction of ${\bf r}-{\bf r}''$ and an
analogous unit vector ${\bf e}'$. The integrand vanishes for all points
${\bf r}''$ that lie on the line connecting ${\bf r}$ and ${\bf r}'$, i.e.
for forward scattering at the spheres, since ${\bf e}=-{\bf e}'$ in that
case. For backward scattering the correction does not vanish.

In second order of the susceptibility one needs an expression for $\langle
f({\bf r})\, f({\bf r}')\rangle$ which takes account of finite-volume
effects. Analogously to \eref{5.1} we write
\begin{eqnarray}
\fl \langle f({\bf r})\, f({\bf r}')\rangle - \langle f({\bf r})\rangle\,
 \langle f({\bf r}')\rangle=
n\int^V d{\bf r}''\, 
\theta(a-|{\bf r}-{\bf r}''|)\, \theta(a-|{\bf r}'-{\bf r}''|)=\nonumber\\
= n\, \theta_V({\bf r}) \int d{\bf r}''\, 
\theta(a-|{\bf r}-{\bf r}''|)\, \theta(a-|{\bf r}'-{\bf r}''|)\nonumber\\
+n\left[\int^V d{\bf r}''\,
\theta(a-|{\bf r}-{\bf r}''|)\, \theta(a-|{\bf r}'-{\bf
r}''|)\right.\nonumber\\
-\left.\theta_V({\bf r}) \int d{\bf r}''\, 
\theta(a-|{\bf r}-{\bf r}''|)\, \theta(a-|{\bf r}'-{\bf r}''|)\right]
\label{5.7}  
\end{eqnarray}
where correlation effects have been omitted, as before. The first term at
the right-hand side leads to a volume contribution. After proper
localization one finds an expression of the same form as the term of order
$[\chi(z)]^2$ in \eref{4.15} with \eref{4.14}, with the integration
extended over $V$.

The remaining terms at the right-hand side of \eref{5.7} vanish when ${\bf
r}$ and/or ${\bf r}'$ are far from the surface. In fact, one may rewrite
$\theta_V({\bf r}'')-\theta_V({\bf r})$ as $[1-\theta_V({\bf
r})]\theta_V({\bf r}'') -\theta_V({\bf r})\, [1-\theta_V({\bf r}'')]$, so
that ${\bf r}$ and ${\bf r}''$ must be on different sides of the
surface. Since the $\theta$-functions in \eref{5.7} imply that these
positions can at most be a distance $a$ apart, they are within a distance
$a$ from the surface. As a consequence, the contribution of the second term
in \eref{5.7} to the average Green function is a surface term. After a
suitable change of variables it can be written as
\begin{eqnarray}
\fl \frac{z^4}{c^4}\, n \, [\chi(z)]^2
\int^S dS'' \int dh'' \int^S dS''' \int dh'''\, 
 \bfsfG_0({\bf r},{\bf r}''_s+h''\, {\bf n}'',z)\cdot\nonumber\\
\fl \cdot\bfsfG_0({\bf r}''_s+h''\, {\bf n}'',{\bf r}'''_s+h'''\, {\bf
n}''',z)\cdot 
\bfsfG_0({\bf r}'''_s+h'''\, {\bf n}''',{\bf r}',z)\, 
F({\bf r}''_s,{\bf r}'''_s,h'',h''')\, .
\label{5.8}
\end{eqnarray}
The function $F$ is defined as
\begin{eqnarray}
\fl F({\bf r}_s,{\bf r}'_s,h,h')=\int dS''\int
dh''\,
\left[-\theta(-h)\, \theta(h''+\half(h+h'))
+\theta(h)\, \theta(-h''-\half(h+h'))\right]\nonumber\\
\fl \times 
\theta\left(a-|{\bf r}_s+\half(h-h')\, {\bf n}-{\bf r}''_s-h''\, {\bf n}|\right)
\theta\left(a-|{\bf r}'_s-\half(h-h')\, {\bf n}-{\bf r}''_s-h''\, {\bf n}|\right)\, .
\label{5.9}
\end{eqnarray}
The normal unit vectors at ${\bf r}_s$, ${\bf r}'_s$ and ${\bf r}''_s$
can be taken identical since these positions are at most a
distance $2a$ apart. For the same reason the second Green function in
\eref{5.8} can be replaced by its short-range approximation:
\begin{equation}
\bfsfG_0({\bf r},0,z)\simeq 
\frac{c^2}{z^2}\, {\cal P} \frac{1}{4\pi r^3}\, \left(\bfsfI-3\frac{{\bf
r}{\bf r}}{r^2}\right)
+\frac{c^2}{3 z^2}\, \delta({\bf r})\, \bfsfI 
\label{5.10}
\end{equation}
as follows from \eref{4.11}.  After substitution of this expression and of
\eref{5.9} in \eref{5.8} the contribution of the delta function in
\eref{5.10} can be evaluated along the same lines as before. One finds the
following on a par with \eref{5.5}:
\begin{equation}
\fl  -\rmi\,\frac{z^2}{30 c^2}\, n\, v_0\, [\chi(z)]^2\, a\, q\int^S dS''\,  
{\bf n}''\cdot ({\bf e}_s+{\bf e}'_s)\, 
\bfsfG_0({\bf r},{\bf r}''_s,z) \cdot 
\bfsfG_0({\bf r}''_s,{\bf r}',z)\, .
\label{5.11}
\end{equation}

The evaluation of the contribution from the dyadic part of \eref{5.10} is
more complicated. Some of the details are given in appendix A. The result is
\begin{eqnarray}
\fl -\rmi\, \frac{z^2}{25 c^2}\, n\, v_0\, [\chi(z)]^2\, a\, q\int^S
dS''\, \bfsfG_0({\bf r},{\bf r}''_s,z)\cdot
(-\twothirds\,\bfsfI\, {\bf n}''\cdot {\bf e}'_s + {\bf e}'_s\, 
{\bf n}'')\cdot \bfsfG_0({\bf r}''_s,{\bf r}',z)\, .\nonumber\\
\label{5.12}
\end{eqnarray}

A further term of second order in $\chi(z)$ arises from the uncorrelated
part $\langle f({\bf r})\rangle\,\langle f({\bf r}')\rangle$ of $\langle
f({\bf r})\, f({\bf r}')\rangle$, as given in \eref{5.7}. Since it is
proportional to $n^2$ it is negligible for a dilute set of scatterers.

The complete set of terms that result from surface effects in second order
of the susceptibility $\chi(z)$ is found by adding \eref{5.11} and
\eref{5.12}:
\begin{eqnarray}
\fl -\rmi\,  \frac{z^2}{25 c^2}\, n\, v_0\, [\chi(z)]^2\, a\, q\int^S
dS''\, \bfsfG_0({\bf r},{\bf r}''_s,z)\cdot
\left[\sixth \bfsfI\, {\bf n}''\cdot (5{\bf e}_s+{\bf e}'_s) 
+{\bf e}'_s\, 
{\bf n}''\right]\cdot \nonumber\\
\cdot\bfsfG_0({\bf r}''_s,{\bf r}',z)\, .
\label{5.13}
\end{eqnarray}
As before we may use Gauss's theorem to write this expression as a volume
integral:
\begin{eqnarray}
\fl -\frac{z^2}{25 c^2}\, n\, v_0\, [\chi(z)]^2\, q^2\int^V d{\bf r}''\, 
\bfsfG_0({\bf r},{\bf r}'',z)\cdot
[\bfsfI\, (1+{\bf e}\cdot{\bf e}') 
+{\bf e}'\,{\bf e}]\cdot \bfsfG_0({\bf r}'',{\bf r}',z)\, .
\label{5.14}
\end{eqnarray}
As in \eref{5.6} the integrand vanishes for forward scattering, since ${\bf
e}=-{\bf e}'$ in that case and $\bfsfG_0({\bf r},{\bf r}'',z)\cdot{\bf
e}=0$ for large $|{\bf r}-{\bf r}''|$.

In conclusion, we have found an expression for the average Green function
$\langle \bfsfG({\bf r},{\bf r}',z)\rangle$ of a dilute set of spherical
scatterers inside a volume $V$. The expression is valid up to first order
in the density and second order in the susceptibility and for positions
${\bf r}$ and ${\bf r}'$ far outside $V$. Its bulk part is given by
\eref{4.15} (with volume integrations extended over $V$), while the
contribution from the surface is the sum of \eref{5.5} (or
\eref{5.6}) and \eref{5.13} (or \eref{5.14}). The complete result is
\begin{eqnarray}
\fl \langle \bfsfG({\bf r},{\bf r}',z)\rangle=\bfsfG_0({\bf r},{\bf
r}',z)\nonumber\\
\fl -\frac{z^2}{c^2}\, n\, v_0\, \chi(z) \int^V d{\bf r}'' \, 
\left[ 1-\fifth\, q^2\, (1+{\bf e}\cdot{\bf e}') \right]
\bfsfG_0({\bf r},{\bf r}'',z) \cdot \bfsfG_0({\bf r}'',{\bf r}',z)\nonumber\\ 
\fl +\frac{z^2}{c^2}\, n\, v_0 \, [\chi(z)]^2
\int^V d{\bf r}''\, \bfsfG_0({\bf r},{\bf r}'',z)\cdot
\left[\bfsfI \left(\frac{1}{3}-\frac{1}{3}\, q^2
-\frac{1}{25}\, q^2\, {\bf e}\cdot{\bf e}'
-\frac{2\rmi}{9}\, q^3\right)
-\frac{1}{25}\, q^2\, {\bf e}'\,{\bf e}\right]\nonumber\\
\cdot\bfsfG_0({\bf r}'',{\bf r}',z)\, .
\label{5.15}
\end{eqnarray}

To check this expression we use it to derive the electric field generated by a
source far away from the volume $V$. The Fourier component ${\bf E}({\bf
r},\omega)$ of the electric field follows from the current density
component ${\bf J}({\bf r},\omega)$ of the source as
\begin{equation}
{\bf E}({\bf r},\omega)=-\rmi\, \mu_0\, \omega \int d{\bf r}' \bfsfG({\bf
r},{\bf r}',\omega+\rmi 0)\cdot {\bf J}({\bf r}',\omega)\, .
\label{5.16}
\end{equation}
If the source is such that in the absence of the medium the electric field
is a plane wave with Fourier component ${\bf E}_i({\bf r},\omega)=E_0\,
{\bf e}_\sigma\, \exp(\rmi {\bf k}\cdot{\bf r})$, with ${\bf e}_\sigma$
being a polarization vector, the average of the full electric-field
component ${\bf E}({\bf r},\omega)$, including the response of the medium,
is obtained from \eref{5.15} as
\begin{eqnarray}
\fl \langle {\bf E}({\bf r},\omega)\rangle={\bf E}_i({\bf r},\omega)\nonumber\\
\fl -\frac{\omega^2}{c^2}\, n\, v_0\, \chi(\omega+\rmi 0) \int^V d{\bf r}'' \, 
\left[ 1-\fifth\, q^2\, (1-{\bf e}\cdot\hat{\bf k}) \right]
\bfsfG_0({\bf r},{\bf r}'',\omega+\rmi 0) \cdot {\bf E}_i({\bf r}'',\omega)\nonumber\\ 
\fl +\frac{\omega^2}{c^2}\, n\, v_0 \, [\chi(\omega+\rmi 0)]^2
\int^V d{\bf r}''\, \bfsfG_0({\bf r},{\bf r}'',\omega+\rmi 0)\cdot\nonumber\\
\cdot\left[\bfsfI \left(\frac{1}{3}-\frac{1}{3}\, q^2
+\frac{1}{25}\, q^2\, {\bf e}\cdot\hat{\bf k}
-\frac{2\rmi}{9}\, q^3\right)
+\frac{1}{25}\, q^2\, \hat{\bf k}\,{\bf e}\right]
\cdot{\bf E}_i({\bf r}'',\omega)\, .
\label{5.17}
\end{eqnarray}
As before, ${\bf e}$ is a unit vector in the direction ${\bf r}-{\bf
r}''$. The unit vector ${\bf e}'$ in \eref{5.15} could be replaced by minus
the unit vector $\hat{\bf k}$ in the direction of the wave vector of the
incoming wave. The expression found here is consistent with that obtained
for the average scattered field from a collection of dielectric spheres in
Mie theory, as is shown in appendix B. 

The above derivation of the average Green function in the presence of a
finite volume filled with absorbing scatterers clearly shows how in general
both bulk and surface effects contribute in producing the complete
result. A naive treatment in which the surface effects are neglected does
not yield the correct answer, when the spheres are of a finite extent. The
surface contributions account for the coarseness of the surface, which
arises from the fact that some of the spheres may protrude.  These
protrusions give rise to specific terms in the average Green function that
do not occur for an infinite medium.

\section{Atomic decay near a half-space of absorptive scatterers}\label{sec6} 

We consider an excited atom in the presence of a medium of absorptive
scatterers that fills the complete half-space $z<0$. The atomic position is
$(0,0,z_a)$, with $z_a>0$. We assume $z_a\, \omega_a/c\gg 1$, so that the
results of the previous sections can be applied. In the electric-dipole
approximation the average decay rate follows from \eref{3.10} by taking the
average over the position of the scatterers:
\begin{equation} 
\langle\Gamma\rangle= -\frac{2\omega_a^2}{\varepsilon_0 \hbarit c^2} 
\,\langle e|\bmu|g\rangle \cdot
{\rm Im}\, \langle\bfsfG({\bf r}_a,{\bf r}_a,\omega_a+\rmi 0)\rangle\cdot 
\langle g|\bmu|e\rangle\, .
\label{6.1}
\end{equation}
At the right-hand side we substitute expression \eref{5.15}
for the average Green function. The leading term yields the standard vacuum
decay rate
\begin{equation}
\Gamma_0=\frac{\omega_a^3}{3\pi \varepsilon_0 \hbarit c^3}\, 
|\langle e|\bmu|g\rangle|^2 \, . 
\label{6.2}
\end{equation}

The next term in \eref{5.15} leads to a first correction in
$\langle\bfsfG({\bf r}_a,{\bf r}_a,\omega_a+\rmi 0)\rangle$ of
the form
\begin{equation}
\fl -\frac{\omega_a^2}{c^2}\, n\, v_0\, \chi(\omega_a+\rmi 0)\,
(1-\twofifths\, q^2) \int^V d{\bf r}\, 
\bfsfG_0({\bf r}_a,{\bf r},\omega_a+\rmi 0)\cdot
\bfsfG_0({\bf r},{\bf r}_a,\omega_a+\rmi 0)
\label{6.3}
\end{equation}
since ${\bf e}={\bf e}'$ in the present case. The integral is a diagonal
tensor, with equal $xx$- and $yy$-components, and a $zz$-component that is
different. For the $xx$- and $yy$-components we find the following upon
substituting the long-range form of the vacuum Green function \eref{4.11}
and using cylinder coordinates:
\begin{eqnarray}
\fl \int_{z_a}^\infty dz\int_0^\infty d\rho\, \rho\, 
\frac{2z^2+\rho^2}{16\pi(z^2+\rho^2)^2}\,
{\rm exp}\left[ 2\rmi\, \frac{\omega_a+\rmi 0}{c}\, 
(z^2+\rho^2)^{1/2} \right]\, .
\label{6.4}
\end{eqnarray}
Introducing the new variable $t=[z^2+\rho^2]^{1/2}/z_a$ instead of
$\rho$ and carrying out the integrals, we get
\begin{equation}
\frac{z_a}{16\pi}\, \left[\fourthirds\, E_0(u)-E_1(u)-\third\, E_3(u)\right]
\label{6.5}
\end{equation}
with $u=-2\rmi\, z_a\, (\omega_a+\rmi 0)/c$ and with the functions 
\begin{equation}
E_n(x)=\int_1^\infty dt\, \frac{\rme^{-xt}}{t^n}
\label{6.6}
\end{equation}
for a non-negative integer $n$ and for $x$ in the right half of the complex
plane.  Since for large $|x|$ these functions are given by their asymptotic
expansions
\begin{equation}
E_n(x)=\frac{\rme^{-x}}{x}\, \left[1-\frac{n}{x}+\frac{n(n+1)}{x^2}+\ldots
\right]
\label{6.7}
\end{equation}
the $xx$- and $yy$-components of the second term in \eref{5.15} read
\begin{equation}
\frac{1}{32\pi}\, n\, v_0\, \chi(\omega_a+\rmi 0)\, 
\left(1-\frac{2}{5}\, q^2\right)
\frac{\rme^{2\rmi z_a \omega_a/c}}{z_a}\, .
\label{6.8}
\end{equation} 
A similar calculation leads to the conclusion that the corresponding
$zz$-component is inversely proportional to $z_a^2$ so that it is small for
large values of $z_a\, \omega_a/c$.

The last term of \eref{5.15} can be evaluated in an analogous way. Upon
using the identity $\bfsfG_0({\bf r},{\bf r}'',\omega_a+\rmi 0)\cdot{\bf
e}= 0$ for large $|{\bf r}-{\bf r}''|\, \omega_a/c$, the $xx$- and
$yy$-components are found as
\begin{equation}
-\frac{1}{32\pi}\, n\, v_0\, [\chi(\omega_a+\rmi 0)]^2\, 
\left( \frac{1}{3}-\frac{28}{75}\, q^2-\frac{2\rmi}{9}\, q^3\right)
\frac{\rme^{2\rmi z_a \omega_a/c}}{z_a}
\label{6.9}
\end{equation} 
while the $zz$-component is inversely proportional to $z_a^2$, as before.

Collecting all results, we have found the following expression for the
average decay rate of an excited atom in the presence of a semi-infinite
medium of absorbing spherical scatterers:
\begin{eqnarray}
\fl \langle \Gamma\rangle =\Gamma_0 
-\frac{3}{16}\,
n\, v_0\, \Gamma_{0,\perp}\nonumber\\
\fl \times {\rm Im}\left\{\left[1-\frac{2}{5}\, q^2 -
\left(\frac{1}{3}-\frac{28}{75}\, q^2-\frac{2\rmi}{9}\, q^3\right)\, 
\chi(\omega_a+\rmi 0)\right]\, \chi(\omega_a+\rmi 0)\, 
\frac{\rme^{2\rmi \zeta_a}}{\zeta_a}\right\}
\label{6.10}
\end{eqnarray}
for large values of the dimensionless distance $\zeta_a=z_a\omega_a/c$
between the atom and the medium. Here $\Gamma_{0,\perp}$ is given by \eref{6.2},
with $\bmu$ replaced by the projection $\bmu_\perp$ of $\bmu$ on the
plane parallel to the interface of the medium.

In deriving the above result we have taken due account of the surface
effects. The terms of order $q^2$ would have been determined incorrectly,
if the contributions of section \ref{sec5} had been missed. As we have
seen, the latter contributions correct for the fact that protruding spheres
cause surface coarseness.

Expression \eref{6.10} is the main result of this paper. It shows the
interplay of absorption and scattering effects in the modification of the
average decay rate. For spheres that do not absorb at the atomic frequency,
so that $\chi(\omega_a)$ is real, the average decay rate can be rewritten
as
\begin{eqnarray}
\fl\langle \Gamma\rangle =\Gamma_0 
-\frac{3}{16}\,
n\, v_0\, \Gamma_{0,\perp}
\, \left\{\left[1-\frac{2}{5}\, q^2
- \left(\frac{1}{3}-\frac{28}{75}\, q^2\right)\,
\chi(\omega_a)\right]\, \chi(\omega_a)\, 
 \frac{\sin(2 \zeta_a)}{\zeta_a}\right.\nonumber\\
\left. +\frac{2}{9}\, q^3\, [\chi(\omega_a)]^2\, 
\frac{\cos(2 \zeta_a)}{\zeta_a}\right\}\, .
\label{6.11}
\end{eqnarray}
On the other hand, if absorption plays a role, whereas scattering effects
can be neglected (as is the case for spheres with $q\approx
0$), one has
\begin{eqnarray}
\fl \langle \Gamma\rangle =\Gamma_0 
-\frac{3}{16}\,
n\, v_0\, \Gamma_{0,\perp}
\, \left[\left\{\chi_r(\omega_a)-\frac{1}{3}[\chi_r(\omega_a)]^2+
\frac{1}{3}[\chi_i(\omega_a)]^2\right\}\, 
 \frac{\sin(2 \zeta_a)}{\zeta_a}\right.\nonumber\\
\left.+\left\{1-\frac{2}{3}\chi_r(\omega_a)\right\}\chi_i(\omega_a)\,    
 \frac{\cos(2 \zeta_a)}{\zeta_a}\right] 
\label{6.12}
\end{eqnarray}
with $\chi_r(\omega)$ and $\chi_i(\omega)$ being the real and imaginary part of
$\chi(\omega+\rmi 0)$, respectively.  The decay rate for an excited atom in
front of an absorbing dielectric half-space without scattering has been
determined before, as we noted in section \ref{sec2}. From the results in
\cite{CCM96,WE99} one derives (after a few minor amendments) a decay rate
for large $\zeta_a$ that coincides with \eref{6.12}, when the effective
susceptibility \eref{4.14} for $q\rightarrow 0$ is introduced.

\begin{figure}
  \begin{center}
    \includegraphics[height=5.cm]{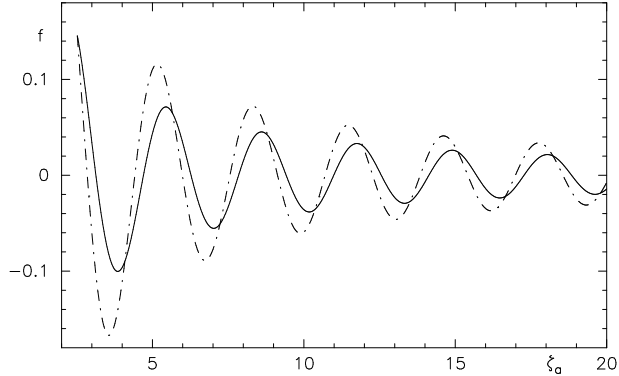}
  \caption{Decay rate correction function
  $f(\zeta_a)=-16\,(\langle\Gamma\rangle-\Gamma_0)/(3\, n\, v_0\,
  \Gamma_{0,\perp})$ for a medium with scattering spheres (with $q=0.5$,
  $\chi(\omega_a)=0.5$, \full) and for a medium with absorbing spheres
  (with $q=0$, $\chi(\omega_a)=0.5+\rmi \, 0.5$, \chain).}
\label{fig1}
  \end{center}
\end{figure}
In figure \ref{fig1} the decay rate correction function, defined as
$f(\zeta_a)=-16\,(\langle\Gamma\rangle-\Gamma_0)/(3\, n\, v_0\,
\Gamma_{0,\perp})$, is given as a function of $\zeta_a$ for two specific
choices of the parameters $q$ and $\chi(\omega_a)$, corresponding to a
purely scattering case, with a decay rate given by \eref{6.11}, and to a
purely absorbing case, with decay rate \eref{6.12}. Both curves show a
characteristic interference pattern. Depending on the precise location of
the atom, the decay rate is either enhanced or reduced with respect to the
vacuum decay rate. The effect is larger for absorbing spheres than for
scattering spheres. Moreover, the phase of the damped oscillations of the two
curves is different. For absorbing spheres the positions of the extrema are
somewhat nearer to the interface than for scattering spheres.

Interference fringes in emission processes due to reflection at an ideal
mirror have been observed experimentally \cite{ERSB01}. The modification of
radiative properties near a dielectric medium has been determined
experimentally a few years ago as well \cite{ICLS04}. It would be
interesting to measure the influence of absorption and scattering in the
medium on the emission processes and to compare the results with those
found here.

\section{Conclusion and outlook}

In this paper we have shown how both scattering and absorption effects can
play a role in the decay of an excited atom in the vicinity of a dielectric
medium. Since atomic decay is essentially a non-classical phenomenon, a
consistent treatment requires the use of a quantum-mechanical description
for the scattering and absorbing dielectric medium, for the atom and for
the electromagnetic fields through which they interact. As we have seen, a
convenient model that suits these requirements is furnished by an
inhomogeneous damped-polariton model for a set of absorptive dielectric
spheres that scatter incoming light. In contrast to what one might expect,
the average bulk properties of such a granular medium are in general not
sufficient to account for all effects of scattering on the atomic decay, at
least for spheres with a finite size as compared to the atomic
wavelength. In fact, under these circumstances the effective dielectric
constant for the bulk does not yield all information on the scattering
processes. A subtle surface effect in the scattering contributes to the
change of the decay as well. Only when this surface contribution is taken
into account does one obtain the complete expression for the modified decay
rate.

In our treatment we have confined ourselves to a description of a dilute
medium in which multiple-scattering effects are negligible. Moreover, we
have considered only the first few terms in an expansion of the scattering
amplitudes with respect to the ratio of the spherical diameter and the
atomic wavelength. It would be interesting to see whether the above
findings about the importance of surface effects hold as well when the
medium gets denser or the spheres bigger. As a further simplification of
our discussion we have assumed that the distance between the atom and the
granular medium is large as compared to the wavelength, so that only the
leading term in a long-range expansion of the decay rate had to be
retained. For smaller distances the analysis of the surface effects gets
more complicated.

\appendix
\section{Surface effects in second order of the susceptibility}\label{apA}

The surface contribution to the average Green function in second order of
the susceptibility has been written in \eref{5.8}. It contains a product of
three vacuum Green functions, the second of which is given by its
short-range approximation \eref{5.10}. The contribution from the delta
function in this Green function has been determined in the main text.  In
this appendix we shall show how the contribution from the dyadic part in
\eref{5.10} can be evaluated.

Substitution of the dyadic part of \eref{5.10} in \eref{5.8} with
\eref{5.9} yields  
the following expression: 
\begin{eqnarray}
\fl \frac{z^2}{4\pi c^2}\, n \, [\chi(z)]^2 \, {\cal P}
\int^S dS'' \int dh'' \int^S dS''' \int dh'''\, 
 \bfsfG_0({\bf r},{\bf r}''_s+h''\, {\bf n}'',z)\cdot\nonumber\\
\fl\cdot \left\{\bfsfI-\frac{3}{|{\bf r}''_s-{\bf r}'''_s|^2+(h''-h''')^2}\, 
\left[({\bf r}''_s-{\bf r}'''_s)\, ({\bf r}''_s-{\bf
r}'''_s)
+(h''-h''')^2\, {\bf n}''\, {\bf n}''\right.\right.\nonumber\\
\fl \left.\left.+(h''-h''')\,  ({\bf r}''_s-{\bf r}'''_s)\, 
{\bf n}''+
(h''-h''')\,{\bf n}''\, ({\bf r}''_s-{\bf r}'''_s)\right]
\rule{0cm}{5mm}\right\}\cdot
\nonumber\\
\fl \cdot\bfsfG_0({\bf r}'''_s+h'''\, {\bf n}'',{\bf r}',z)\, 
\bar{F}({\bf r}''_s,{\bf r}'''_s,h'',h''')\, .
\label{A.1}
\end{eqnarray}
The surface elements $dS''$ and $dS'''$ are located at ${\bf r}''_s$
and ${\bf r}'''_s$, respectively. The normal unit vectors at these two
positions are almost equal and have been denoted by ${\bf n}''$. The
principal value sign indicates the exclusion of a small sphere around ${\bf
r}''_s+h''\, {\bf n}''$ in the integrations over ${\bf r}'''_s$ and
$h'''$. Furthermore, $\bar{F}({\bf r}_s,{\bf r}'_s,h,h')$ stands for 
$F({\bf r}_s,{\bf r}'_s,h,h')/[|{\bf r}_s-{\bf r}'_s|^2+
(h-h')^2]^{3/2}$, while the variable $z$ equals $\omega+\rmi 0$, as before.

As in \eref{5.4}, the two Green functions in \eref{A.1} can be expanded (in
their second or first argument, respectively) around their values at ${\bf
r}''_s$, if both $|{\bf r}-{\bf r}''_s|$ and $|{\bf r}'-{\bf
r}''_s|$ are large compared to the wavelength, as we have assumed
before. Up to first order in $\omega a/c$, the ensuing phase factor can be 
expanded as
\begin{equation}
1 -\rmi \, \frac{z}{c}\, h''\, {\bf  n}''\cdot{\bf e}_s
-\rmi \, \frac{z}{c}\, h'''\, {\bf  n}''\cdot{\bf e}'_s
+\rmi \, \frac{z}{c}\, ({\bf r}''_s-{\bf r}'''_s)\cdot{\bf
e}'_s
\label{A.2}
\end{equation}
since $h''$, $h'''$ and $|{\bf r}''_s-{\bf r}'''_s|$ are all of order $a$
at most. As before, ${\bf e}_s$ and ${\bf e}'_s$ are unit vectors in the
direction of ${\bf r}-{\bf r}''_s$ and ${\bf r}'-{\bf r}''_s$,
respectively.  The product of \eref{A.2} and the expression between curly
brackets in \eref{A.1} contains all information on the dependence of the
integrand on ${\bf r}''_s-{\bf r}'''_s$. This product may be replaced by
the effectively equivalent form
\begin{eqnarray}
\fl  
\left\{\left[-\half\, |{\bf r}''_s-{\bf r}'''_s|^2+
(h''-h''')^2\right]\, \left(1-
\rmi\, \frac{z}{c}\, h''\, {\bf n}''\cdot{\bf e}_s
-\rmi\, \frac{z}{c}\, h'''\, {\bf n}''\cdot{\bf e}'_s\right)
\, (\bfsfI-3\, {\bf n}''{\bf n}'')\right.\nonumber\\
\fl \left. -\threehalves\, \rmi \, \frac{z}{c}\, (h''-h''')\, 
|{\bf r}''_s-{\bf r}'''_s|^2\, 
\left({\bf e}'_s\, {\bf n}''+{\bf n}''\,{\bf e}'_s
-2\, {\bf n}''{\bf n}''\, {\bf n}''\cdot{\bf e}'_s\right)\right\}  
\nonumber\\
\times[|{\bf r}''_s-{\bf r}'''_s|^2+(h''-h''')^2]^{-1}
\label{A.3}
\end{eqnarray}
up to first order in $\omega a/c$, when use is made of the rotation
symmetry of the integration over ${\bf r}'''_s$ in the planar surface
through ${\bf r}''_s$ and orthogonal to ${\bf n}''$. This symmetry implies 
that the product $({\bf r}''_s-{\bf r}'''_s)\,({\bf r}''_s-{\bf r}'''_s)$
may be replaced by $\half\, |{\bf r}''_s-{\bf r}'''_s|^2\, 
(\bfsfI-{\bf n}''{\bf n}'')$.   

To simplify the multiple integral in \eref{A.1} we introduce, instead of
$h''$ and $h'''$, their sum and difference as new integration
variables. Subsequently, we carry out the integral over $h''+h'''$, at
fixed $h''-h'''$.  Inspecting the result, one finds that in leading order
of $\omega a/c$ the expression \eref{A.1} vanishes owing to the odd parity
of the integrand in its variable $h''-h'''$. The contribution from the next
order in $\omega a/c$ does not vanish. It can be determined by first
considering the integrations that are hidden in the definition of the
function $F$, as given in \eref{5.9}. These lead to the following
integrals:
\begin{eqnarray}
\fl I_1({\bf r}_s,h)=\int^S dS'\int dh'\, 
\theta[a^2-|\half\, {\bf r}_s-{\bf r}'_s|^2-(\half\, h-h')^2]
\nonumber\\
\times\theta[a^2-|\half\, {\bf r}_s+{\bf r}'_s|^2-(\half\, h+h')^2]
\label{A.4}\\
\fl I_2({\bf r}_s,h)=\int^S dS'\int dh'\, h'^2\,  
\theta[a^2-|\half\, {\bf r}_s-{\bf r}'_s|^2-(\half\, h-h')^2]
\nonumber\\
\times\theta[a^2-|\half\, {\bf r}_s+{\bf r}'_s|^2-(\half\, h+h')^2]
\label{A.5}
\end{eqnarray}
with $h>0$. In writing these integrals we have chosen the origin of the
coordinate system to be situated at the surface.  Both of the integrals
vanish for $r\equiv [|{\bf r}_s|^2+h^2]^{1/2} \geq 2\, a$. 

In terms of the above integrals the contribution \eref{A.1} to the
average Green function becomes in leading order of $\omega a/c$:
\begin{eqnarray}
\fl -\rmi \frac{z^3}{4\pi c^3}\, n\, [\chi(z)]^2\, {\cal P}\int^S dS''
\int^S dS'''\int dh'''\, [|{\bf r}''_s-{\bf
r}'''_s|^2+(h''-h''')^2]^{-5/2}\, 
\bfsfG_0({\bf r},{\bf r}''_s,z)\cdot\nonumber\\
\fl\cdot \left[ [-\half \, |{\bf r}''_s-{\bf r}'''_s|^2+(h''-h''')^2]
\left\{ [\half I_2-\eighth (h''-h''')^2\, I_1]\, 
{\bf n}''\cdot({\bf e}_s +{\bf e}'_s)\right. \right. \nonumber\\
\fl \left. +\fourth (h''-h''')^2\, I_1\, 
{\bf n}''\cdot ({\bf e}_s-{\bf e}'_s)\right\}
(\bfsfI-3\, {\bf n}''{\bf n}'')\nonumber\\
\fl \left. +\threefourth\, (h''-h''')^2\, |{\bf r}''_s-{\bf r}'''_s|^2\, I_1\, 
({\bf e}'_s\, {\bf n}''+{\bf n}''\, {\bf e}'_s
-2\, {\bf n}''\, {\bf n}''\, {\bf n}''\cdot{\bf e}'_s)\right]\cdot
\bfsfG_0({\bf r}''_s,{\bf r}',z)
\label{A.6}
\end{eqnarray}
with $I_1$ and $I_2$ depending on ${\bf r}''_s-{\bf r}'''_s$ and
$h''-h'''$. 

In order to proceed we have to determine explicit expressions for the
integrals $I_1$ and $I_2$. The first is the standard overlap integral,
which has been encountered in \eref{4.4}. It equals $v_0\, c(r)$, with
$c(r)$ given in \eref{4.7}. The integral $I_2$ is the second moment of the
overlap integral. Choosing cartesian coordinates in such a way that the
normal to the surface at the origin points in the direction of the positive
$z$-axis, and that ${\bf r}_s$ equals $(x,0,0)$ with $x>0$, we may write
\eref{A.5} as
\begin{eqnarray}
\fl I_2({\bf r}_s,h)=\int_{-\infty}^{\infty}dx'\int_{-\infty}^{\infty}dy'
\int_{-\infty}^{\infty}dh'\, h'^2\, 
\theta\left[a^2-(\half x-x')^2-y'^2-(\half h-h')^2\right]\nonumber\\
\times\theta\left[a^2-(\half x+x')^2-y'^2-(\half h+h')^2\right]\, .
\label{A.7}
\end{eqnarray}
The integral over $y'$ is trivial, with the result
\begin{eqnarray}
\fl I_2({\bf r}_s,h)=4\int_{-\infty}^\infty dh'\, h'^2\int_{-hh'/x}^{\infty}dx'\, 
\left[a^2-(\half x+x')^2-(\half h+h')^2\right]^{1/2}\nonumber\\
\times\theta\left[a^2-(\half x+x')^2-(\half h+h')^2\right]\, .
\label{A.8}
\end{eqnarray}

The $\theta$-function constrains the integrations over $x'$ and $h'$. This
constraint interferes with the bounds on the integration written explicitly
in \eref{A.8}. By a geometrical analysis one finds that the interference
depends on the sign of the combination $x^2+h^2-2\, a\, h$. In fact, for
$2\, a\, h \leq x^2+h^2\leq 4\, a^2$ the upper limit
of the $x'$-integration is effectively finite, while the lower limit is
left unchanged. Furthermore, the domain of the $h'$-integral is found to be
constrained to values $|h'|\leq M$, with $M=x\, [(a^2-\fourth\,
x^2-\fourth\, h^2)/(x^2+h^2)]^{1/2}$. Hence, after a shift of the
$x'$-variable one has the following for $2\, a\, h \leq x^2+h^2\leq 4\, a^2$:
\begin{eqnarray}
\fl I_2({\bf r}_s,h)=4\int_{-M}^M dh'\, h'^2\int^{[a^2-(h/2+h')^2]^{1/2}}
_{-hh'/x+x/2}dx'\, \left[a^2- x'^2-(\half h+h')^2\right]^{1/2}\, .
\label{A.9}
\end{eqnarray}
The integral over $x'$ yields the result
\begin{eqnarray}
\fl \fourth\, \pi \left[a^2-(\half\, h+h')^2\right] -
\half\left(\half \, x-\frac{h\, h'}{x}\right)\, 
\left[ a^2-\fourth\, x^2-\fourth\,
h^2-h'^2\,\frac{x^2+h^2}{x^2}\right]^{1/2}\nonumber\\
\fl -\half\, \left[a^2-(\half \, h+h')^2\right]\, \arcsin 
\left[\left.\left(\half \, x-\frac{h\, h'}{x}\right) \right/
\left[a^2-(\half \, h+h')^2\right]^{1/2}\right]\, .
\label{A.10}
\end{eqnarray}
After a partial integration in order to get rid of the arcsine function, the
integral over $h'$ can be carried out as well. The final result is
\begin{eqnarray}
\fl I_2({\bf r}_s,h)=\frac{\pi}{\sqrt{x^2+h^2}}\, 
\left[ a^4\, \left(-\fourth\, x^2-\half\, h^2\right)+
a^2\, \left(\textstyle{\frac{1}{24}}\, x^4-
\textstyle{\frac{1}{24}}\, x^2\, h^2 -
\textstyle{\frac{1}{12}}\, h^4\right)\right.\nonumber\\
\fl \left. -\textstyle{\frac{1}{320}}\, x^6
-\textstyle{\frac{1}{240}}\, x^4\, h^2 
+\textstyle{\frac{1}{960}}\, x^2\, h^4
+\textstyle{\frac{1}{480}}\, h^6\right]
+ \textstyle{\frac{1}{15}}\, \pi \,  a^3\, \left(4\, a^2+5\, h^2\right)
\label{A.11}
\end{eqnarray}
for $2\, a\, h \leq x^2+h^2\leq 4\, a^2$.

In the other case $x^2+h^2 < 2\, a\, h$ the double integral is the sum of
two contributions with different bounds:
\begin{eqnarray}
\fl I_2({\bf r}_s,h)=4\int_{-M}^M dh'\, h'^2\int^{[a^2-(h/2+h')^2]^{1/2}}
_{-hh'/x+x/2}dx'\, \left[a^2-x'^2-(\half
h+h')^2\right]^{1/2}\nonumber\\
\fl +\, 4\int_{M}^{-h/2+a} dh'\, h'^2\int^{[a^2-(h/2+h')^2]^{1/2}}
_{-[a^2-(h/2+h')^2]^{1/2}}dx'\, \left[a^2- x'^2-(\half
h+h')^2\right]^{1/2}\, .
\label{A.12}
\end{eqnarray}
Upon evaluating the $x'$- and the $h'$-integral we arrive at a result that
is found to be identical to that given in \eref{A.11}.

Employing spherical coordinates in \eref{A.11}, with $h=r\, \cos\theta$ and
$x =r\, \sin\theta$, we may write the second moment of the
overlap integral for both cases as
\begin{eqnarray}
\fl I_2({\bf r}_s,h)=v_0\, a^2\, \left[\frac{1}{5}-\frac{3r}{16a}\,
(1+\cos^2\theta)+\frac{r^2}{4a^2}\, \cos^2\theta+\frac{r^3}{32a^3}\, 
(1-3\, \cos^2\theta)\right.\nonumber\\
\left. -\frac{r^5}{1280 a^5}\, (3-5\, \cos^2\theta)\right]\,
\theta(2\, a-r)\, .
\label{A.13}
\end{eqnarray}

Having obtained explicit expressions for $I_1$ and $I_2$, we return to
\eref{A.6}.  The integrals over ${\bf r}'''_s$ (with surface element
$dS'''$) and $h'''$ can be calculated straightforwardly upon introducing
spherical coordinates and performing the angular integration first. In this
way we arrive at the result
\begin{eqnarray}
\fl -\rmi\, \frac{z^2}{25 c^2}\, n\, v_0\, [\chi(z)]^2\, a\, q\int^S
dS''\, \bfsfG_0({\bf r},{\bf r}''_s,z)\cdot
(-\twothirds\,\bfsfI\, {\bf n}''\cdot {\bf e}'_s + {\bf e}'_s\, 
{\bf n}''+{\bf n}''\, {\bf e}'_s)\cdot\nonumber\\
\cdot \bfsfG_0({\bf r}''_s,{\bf r}',z)\, .
\label{A.14}
\end{eqnarray}
This is equivalent to \eref{5.12}, since one may use the identity ${\bf
e}'_s\cdot\bfsfG({\bf r}''_s,{\bf r}',z)=0$ for distances $|{\bf
r}'-{\bf r}''_s|$ that are large compared to the wavelength. 

\section{Scattering fields in Mie theory}\label{apB}

Electromagnetic scattering from a dielectric sphere has first been treated
by Mie \cite{M08} and reviewed subsequently by several authors
\cite{H81,BW99}. If a linearly polarized incoming plane wave, with the wave
vector ${\bf k}$ in the direction of the positive $z$-axis, with the 
polarization vector ${\bf e}_\sigma$ along the $x$-axis and with the amplitude
$E_0$, impinges on a dielectric sphere with a radius $a$, with centre at the
origin and with a dielectric constant $\varepsilon=1+\chi$, the components
of the scattered electric field in the far-field region have the form
\cite{BW99}
\begin{eqnarray} 
\fl E^{(s)}_\theta(r,\theta,\varphi)=E_0\, \frac{\rme^{\rmi k r}}{kr}\,\cos\varphi
\sum_{\ell=1}^\infty(-\rmi)^\ell\, \left[ B^e_\ell\, \tau_\ell(\cos
\theta)+B^m_\ell\, \pi_\ell(\cos\theta)\right] \nonumber\\ 
\fl E^{(s)}_\varphi(r,\theta,\varphi)=-E_0\, \frac{\rme^{\rmi k r}}{kr}\,\sin\varphi
\sum_{\ell=1}^\infty(-\rmi)^\ell\, \left[ B^e_\ell\, \pi_\ell(\cos
\theta)+B^m_\ell\, \tau_\ell(\cos\theta)\right] 
\label{B.1} 
\end{eqnarray}
with spherical coordinates $r$, $\theta$ and $\varphi$. The angular functions
are defined in terms of associated Legendre polynomials as 
\begin{equation}
\fl \pi_\ell(\cos\theta)=\frac{1}{\sin\theta}\, P^1_\ell(\cos\theta)\quad ,
\quad \tau_\ell(\cos\theta)=\frac{d}{d\theta}\, P^1_\ell(\cos\theta)\, .
\label{B.2}
\end{equation}
The electric and magnetic multipole amplitudes read
\begin{equation}
B^p_\ell=\rmi^{\ell+1}\, \frac{2\ell+1}{\ell(\ell+1)}\, \frac{N^p_\ell}{D^p_\ell}
 \label{B.3}
\end{equation}
with $p=e,m$. The numerators and denominators are given as
\begin{eqnarray}
\fl N^e_\ell=\varepsilon\, \left[(\ell+1)\, j_\ell(q)-q\,
j_{\ell+1}(q)\right]\, j_\ell(q')-
\left[(\ell+1)\, j_\ell(q')-q'\,
j_{\ell+1}(q')\right]\, j_\ell(q) \nonumber\\
\fl N^m_\ell=\left[(\ell+1)\, j_\ell(q)-q\,
j_{\ell+1}(q)\right]\, j_\ell(q')-
\left[(\ell+1)\, j_\ell(q')-q'\,
j_{\ell+1}(q')\right]\, j_\ell(q) \nonumber\\
\fl D^e_\ell=\varepsilon\, \left[(\ell+1)\, h^{(1)}_\ell(q)-q\,
h^{(1)}_{\ell+1}(q)\right]\, j_\ell(q')-
\left[(\ell+1)\, j_\ell(q')-q'\,
j_{\ell+1}(q')\right]\, h^{(1)}_\ell(q) \nonumber\\
\fl D^m_\ell=\left[(\ell+1)\, h^{(1)}_\ell(q)-q\,
h^{(1)}_{\ell+1}(q)\right]\, j_\ell(q')-
\left[(\ell+1)\, j_\ell(q')-q'\,
j_{\ell+1}(q')\right]\, h^{(1)}_\ell(q) 
\label{B.4}
\end{eqnarray}
with spherical Bessel and Hankel functions depending on $q=k a$ and
$q'=\sqrt{\varepsilon}\, q$.

For small values of $q$, the first few multipole amplitudes  get the form
\begin{eqnarray}
B^e_1=\rmi\, q^3\, \frac{\chi}{3+\chi}\, \left( 1-\frac{3}{5}\, q^2\, 
\frac{1-\chi}{3+\chi} +\frac{2\rmi}{3}\, q^3\,
\frac{\chi}{3+\chi}\right) \nonumber\\
B^e_2=-\frac{1}{18}\, q^5\, \frac{\chi}{5+2\chi} \nonumber\\
B^m_1=\frac{\rmi}{30}\, q^5\, \chi  \label{B.5}
\end{eqnarray}
up to the order $q^6$. When $\chi$ is small as well, the first two of these can be
written as 
\begin{eqnarray}
B^e_1=\frac{\rmi}{3}\, q^3\, \chi\, \left[
1-\frac{1}{3}\, \chi-\frac{1}{5}\, q^2\, \left(1-\frac{5}{3}\, \chi\right)
+\frac{2\rmi}{9}\, q^3\, \chi\right] \nonumber\\
B^e_2=-\frac{1}{90}\, q^5\, \chi \left(1-\frac{2}{5}\, \chi\right)
\label{B.6} 
\end{eqnarray}
up to the order $\chi^2$. Upon substitution in \eref{B.1} the far fields are found as
\begin{eqnarray}
\fl E^{(s)}_\theta(r,\theta,\varphi)
=E_0\, \frac{\rme^{\rmi k r}}{kr}\, q^3\,\chi\,  \cos\varphi\, 
\left\{\frac{1}{3}\, \left(1-\frac{1}{5}\, q^2\right)\, \cos\theta
+\frac{1}{15}\, q^2\, \cos^2\theta\right.\nonumber\\
\fl\left. +\chi\, \left[-\frac{1}{9}\left(1- q^2-\frac{2\rmi}{3}
\, q^3\right)\, \cos\theta -\frac{1}{75}\, q^2\, 
\left(2\, \cos^2\theta-1\right)\right]\right\}  \nonumber\\
\fl E^{(s)}_\varphi(r,\theta,\varphi)
=-E_0\, \frac{\rme^{\rmi k r}}{kr}\,q^3\, \chi\, \sin\varphi \, 
\left\{\frac{1}{3}\, \left(1-\frac{1}{5}\, q^2\right)
+\frac{1}{15}\, q^2\, \cos\theta\right.\nonumber\\
\fl\left. +\chi\, \left[-\frac{1}{9}\left(1- q^2-\frac{2\rmi}{3}
\, q^3\right) -\frac{1}{75}\, q^2\, \cos\theta\right]\right\} 
\label{B.7}
\end{eqnarray}
up to the order $q^6$ and $\chi^2$. In vectorial notation this expression may be
rewritten by introducing the long-range form of the vacuum Green function
\eref{4.11}:
\begin{eqnarray}
\fl {\bf E}^{(s)}({\bf r})=-\frac{\omega^2}{c^2}\, v_0 \, \chi
\, \bfsfG_0({\bf r},0,\omega+\rmi 0)\cdot
\left\{\bfsfI\, \left(1-\frac{1}{5}\, q^2+\frac{1}{5}\, q^2\, 
\hat{\bf r}\cdot \hat{\bf k}\right)\right.\nonumber\\
\fl\left. +\chi\, \left[\bfsfI\, \left(-\frac{1}{3}+\frac{1}{3}\, q^2
+\frac{2\rmi}{9}\, q^3\right)-\frac{1}{25}q^2\left(
\bfsfI\, \hat{\bf r}\cdot\hat{\bf k}+\hat{\bf k}\, \hat{\bf r}\right)\right]
\right\}\cdot{\bf E}_i(0) \label{B.8}
\end{eqnarray} 
with the spherical volume $v_0=4\, \pi\, a^3/3$ and with the unit vectors
$\hat{\bf r}={\bf r}/r$ and $\hat{\bf k}={\bf k}/k$. This expression for
the scattered electric field is consistent with that found in \eref{5.17}
for the average field due to scattering from a set of Mie spheres.


\section*{References}
\begin{thebibliography}{20}

\bibitem{P46} Purcell E M 1946 {\em Phys.~Rev.} {\bf 69} 681

\bibitem{D70} Dissado L A 1970 {\em J.~Phys.~C} {\bf 3} 94

\bibitem{NA76} Nienhuis G and Alkemade C Th J 1976 {\em Physica~C} {\bf 81} 181

\bibitem{GL91} Glauber R J and Lewenstein M 1991 {\em Phys.~Rev.~A} {\bf
43} 467

\bibitem{M95} Milonni P W 1995 {\em J.~Mod.~Opt.} {\bf 42} 1991

\bibitem{BHL92} Barnett S M, Huttner B and Loudon R 1992 {\em
  Phys.~Rev.~Lett.} {\bf 68} 3698

\bibitem{HK93} Ho S-T and Kumar P 1993 {\em J.~Opt.~Soc.~Am.~ B} {\bf 10} 1620 

\bibitem{J95} Juzeliunas G 1995 {\em Chem.~Phys.} {\bf 198} 145 

\bibitem{BHLM96} Barnett S M, Huttner B, Loudon R and Matloob R 1996 {\em
  J.~Phys.~B} {\bf 29} 3763

\bibitem{J97} Juzeliunas G 1997 {\em Phys.~Rev.~A} {\bf 55} 4015 

\bibitem{SKWB99} Scheel S, Kn\"{o}ll L, Welsch D-G and Barnett S M 1999
{\em Phys.~Rev.~A} {\bf 60} 1590

\bibitem{F99} Fleischhauer M 1999 {\em Phys.~Rev.~A} {\bf 60} 2534

\bibitem{SKW99a} Scheel S, Kn\"{o}ll L and Welsch D-G 1999 {\em Phys.~Rev.~A}
{\bf 60} 4094, errata 2000 {\em Phys.~Rev.~A} {\bf 61} 069901

\bibitem{DKW00} Dung H T, Kn\"{o}ll L and Welsch D-G 2000 {\em
Phys.~Rev.~A} {\bf 62} 053804 

\bibitem{A75} Agarwal G S 1975 {\em Phys.~Rev.~A} {\bf 12} 1475 

\bibitem{KL91} Khosravi H and Loudon R 1991 {\em Proc.~Roy.~Soc.~A} {\bf
433} 337  

\bibitem{CCM96} Courtois J-Y, Courty J-M and Mertz J C 1996 {\em
Phys.~Rev.~A} {\bf 53} 1862

\bibitem{YG96} Yeung M S and Gustafson T K 1996	{\em Phys.~Rev.~A} {\bf 54}
5227 

\bibitem{WE99} Wu S-T and Eberlein C 1999 {\em Proc.~Roy.~Soc.~A} {\bf 455}
2487 

\bibitem{SKW99b} Scheel S, Kn\"{o}ll L and Welsch D-G 1999 {\em Acta
Phys.~Slov.} {\bf 49} 585 

\bibitem{KSW01} Kn\"{o}ll L, Scheel S and Welsch D-G 2001 {\em Coherence
and Statistics of Photons and Atoms}, ed J~Pe\v{r}ina (New York: Wiley) p
1

\bibitem{ICLS04} Ivanov V V, Cornelussen R A, van Linden van den Heuvell H
B and Spreeuw R J C 2004 {\em J.~Opt.~B} {\bf 6} 454 

\bibitem{YV2007} Yannopapas V and Vitanov N V 2007 {\em Phys.~Rev.~B} {\bf
75} 115124

\bibitem{HB92b}Huttner B and Barnett S M 1992 {\em Phys.~Rev.~A} {\bf 46}
4306

\bibitem{SWo04} Suttorp L G and van Wonderen A J 2004 {\em Europhys.~Lett.}
{\bf 67} 766

\bibitem{F61} Fano U 1961 {\em Phys.~Rev.} {\bf 124} 1866

\bibitem{W89} Weiglhofer W 1989 {\em Am.~J.~Phys.} {\bf 57} 455

\bibitem{VCL98} de Vries P, van Coevorden D V and Lagendijk A 1998 {\em
Rev.~Mod.~Phys.} {\bf 70} 447

\bibitem{ERSB01} Eschner J, Raab Ch, Schmidt-Kaler F and Blatt R 2001 {\em
Nature} {\bf 413} 495

\bibitem{M08} Mie G 1908 {\em Ann.~Physik} {\bf 25} 377

\bibitem{H81} van de Hulst H C 1981 {\em Light Scattering by Small
Particles} (New York: Dover)

\bibitem{BW99} Born M and Wolf E 1999 {\em Principles of Optics}
(Cambridge: Cambridge University Press) (section 14.5)



\end{thebibliography}
\end{document}